\documentclass[preprint,aps,superscriptaddress,nofootinbib,
tightenlines,floats,floatfix]{revtex4}

\newcommand{\beq}{\begin{equation}}
\newcommand{\eeq}{\end{equation}}
\newcommand{\beqa}{\begin{eqnarray}}
\newcommand{\eeqa}{\end{eqnarray}}

\newcommand{\ov}{\overline}

\begin{document}

\preprint{\vbox{
  \hbox{SCIPP--2004/20} 
}}

\title{\boldmath Grand Unification with Higher Rank Product Groups }

\vspace*{1.5cm}

\author{Erik Kramer}
\affiliation{Santa Cruz Institute for Particle Physics\\
     Santa Cruz CA 95064, USA\\
lunenor@physics.ucsc.edu\\
$\phantom{}$}

\begin{abstract}

Various ideas support the notion that the GUT gauge group might be a semi-simple direct-product group
such as $SU(5) \times SU(5)$. 
The doublet-triplet splitting problem can be solved  with a 
direct product group. String theory suggests that the 
GUT scale is a modulus. Requiring this rules out a single $SU(5)$ gauge group. 
A model with $SU(5) \times SU(5)$ gauge symmetry and the GUT
scale as a modulus has been shown to exist.
It is shown that extending these ideas to $SO(10) \times SO(10)$
cannot be done with the above requirement without unwanted massless modes at lower energy scales that spoil
the unification of couplings. Therefore these two conditions highly constrain the class of possible GUT models.
\end{abstract}

\maketitle

\section{Introduction}
The idea of unifying the gauge interactions of the Standard Model in one gauge group has long been
appealing to theorists, both for the aesthetic virtue of explaining Standard Model physics under one
simple gauge group, and for the experimental predictions such unifications make, e.g. ${\rm sin} \theta_{\rm W}$.
One of the problems with implementing GUTs is keeping massless the color triplet partners to the Higgs
doublet. As in \cite{Witten, Barr} ,
one way to solve this problem is to introduce a second copy of
the fundamental group and then introduce a discrete symmetry. Breaking the unified group can leave a
combination of this symmetry and a gauge transformation unbroken, and this symmetry forbids 
masses for doublets while allowing triplet mass terms. Witten showed that this cannot be accomplished with a single group. 
Witten considered GUTs motivated by deconstruction of higher dimensional models. 
One can consider taking the extra dimension to be a lattice, rather than a continuum, of points.
In particular, 
consider a lattice of only two points, with the fundamental groups on each
point. This picture may possibly be generalized, but we restrict ourselves to the simplest picture. 

Dine et al. point out in \cite{dnd} that if one considers the construction of grand unified models in string theory
that the fields required to break unification are approximately or exactly massless. The GUT scale
is about two orders of magnitude below the Planck scale. 
Some adjoints will be massless at the Planck scale, and acquire very large VEVs at the GUT scale.
These fields must have very flat potentials in order for this to happen.
Even without looking at GUT models with a string theory bias, an approximately flat potential
gives a natural way of obtaining the GUT scale from the Planck scale and the supersymmetry breaking scale,
as we will argue. With models we describe we will not give any explanation of the relative values
of these scales, however, we will outline how the ratio of scales might arise naturally. 

We will show as well that flat directions in the GUT potential are difficult to realize with a single $SU(5)$
gauge group. Although we already know that we need more than just a simple gauge group to solve
the problem of doublets and triplets, it is interesting
to see that this is not the only reason to consider
using semi-simple gauge groups.

We find that if we require a theory to have approximate flat directions (exact up to nonrenormalizable
terms), forbid the input of explicit mass terms, and implement Witten's ideas for solving the doublet triplet splitting problem
with only discrete symmetries, that if our fundamental gauge group is $SO(10) \times SO(10)$ this cannot be 
done successfully. The fact that extending these ideas to the next simplest group is not possible suggests
that the result with an $SU(5) \times SU(5)$ gauge group is unique in this capacity. Although not all possible 
gauge groups are eliminated at this juncture, it is quite clear that there exists a set of criterion that cannot be met 
by $SO(10) \times SO(10)$ and groups of which this is a subgroup. 
Thus, the criteria we described above are extremely selective.

\label{intro}

\section{Motivations}
\label{doublettriplet}
An important feature of our models should be that the Higgs doublet does not acquire mass at
the GUT scale, while the Higgs triplet acquires a large mass to suppress proton
decay. As discussed by Witten \cite{Witten}, an unbroken discrete symmetry that is not
a subgroup of hypercharge, under which the Higgs doublets and triplets transform differently, can
explain the existence of massless doublets with massive triplets.

As shown by Witten, this is not possible with a single $SU(5)$, $SO(10)$ 
or $E_6$. Rather, he suggests the use of a semi-simple group, such as $SU(5) \times SU(5)$,
and shows that in such models it is possible to use a discrete symmetry to allow triplet
mass terms while forbidding doublet masses at the GUT scale. Let us review Witten's argument.
In a model with a single $SU(5)$, one might imagine putting the Higgs doublets
in chiral superfields $H$ and $\bar{H}$ that transform as a $5$ and a $\bar{5}$,
respectively. These fields then contain color triplets $q$ and $\bar{q}$, and the 
doublets $h$ and $\bar{h}$. The triplets have couplings related by $SU(5)$
to the couplings required of the doublets to give mass to quarks and leptons. 
The couplings of the triplets mediate proton decay, and must therefore have masses close
to the GUT scale for the proton lifetime to be long enough \cite{Witten}. Introduce
a discrete symmetry under which the $(q, h)$ transform as $(e^{i \alpha} , e^{i \beta})$
and the $(\bar{q}, \bar{h})$ transform as $(e^{i \tilde{\alpha}} , e^{i \tilde{\beta}})$. For 
$e^{i(\alpha + \tilde{\alpha})} = 1$ and $e^{i(\beta + \tilde{\beta})} \neq 1$
the doublet mass term is not allowed, and the triplets are allowed to gain mass. 
Hopefully, at the supersymmetry breaking scale this discrete symmetry is 
spontaneously broken so that the Higgs doublets can gain the appropriate mass
for phenomenology. 

Up to a gauge transformation, in $SU(5)$ a discrete symmetry of the low
energy theory commutes with $SU(5)$ \cite{Witten}. For example, this symmetry
might be a combination of a discrete hypercharge transformation, and a discrete
symmetry that commutes with $SU(5)$. Both doublet and triplet mass
terms are invariant under gauge transformations, and a symmetry that commutes 
with $SU(5)$ will transform both mass terms in the same way, so they are either 
both forbidden or both allowed. By introducing a product gauge group, $SU(5) \times SU(5)$
for example,  it is possible
to have the doublet and triplet mass terms transform differently under a
discrete symmetry, in particular by having the superfields that contain the
higgs doublet and color triplet fields transform under different $SU(5)$'s. 

It has been shown by Dine, Nir,
and Shadmi \cite{dnd} that it is possible with the group $SU(5) \times SU(5)$ and discrete symmetries
to construct  models that solve the doublet triplet splitting problem and have an approximately flat 
potential in the GUT breaking direction, and furthermore have no extra massless particles that might spoil
the prediction of coupling constant unification. Models with exact or approximate flat directions in the symmetry breaking
potential have the desireable feature that the value of $M_{\rm GUT}$ need not be a fundamental scale, but rather
can arise dynamically.
As discussed in \cite{dnd}, models with approximate flat directions are those for which the symmetries
forbid renormalizable operators.
Suppose that the lowest dimensional operator in the superpotential contributing to 
the F-term potential is
\beq\label{noperator}
W = {1 \over M^{n-3}_{\rm PL}} X^n
\eeq
and that once supersymmetry is broken a small negative mass squared is generated, giving rise to a potential
of the form
\beq\label{negmass}
V = -m^2 | X |^2 + {1 \over M^{2n-6}_{\rm Pl}} | X |^{2n-2}.
\eeq
This gives a VEV
\beq
\langle X \rangle \sim \left ( m \over M_{\rm Pl} \right )^{1 \over {n-2}} M_{\rm Pl}.
\eeq
If $m$ is at the weak scale, $n$ around 10 will give a VEV on the order of $M_{\rm GUT}$. In models with
approximate flatness we therefore hope to have exact flatness at the level of renormalizable terms,
and to have the lowest allowable nonrenormalizable terms be of mass dimension 10.

In the case of a model where exact flat directions can be achieved, the value of $M_{\rm GUT}$ is fixed 
by supersymmetry breaking, or possibly by some other mechanism.


\section{Single Gauge Groups}

The use of product groups is further motivated by the difficulty of obtaining flat theories from a single
gauge group. The problem lies in that we want a flat potential, but don't want massless modes
below the GUT scale that will spoil the unification of couplings. Therefore, we need to 
generate mass-terms with a non-trivial potential and VEVs. Ideally, the potential we write 
would be one with terms that are allowed by some discrete symmetries or discrete R-symmetries. 
Consider $SU(5)$ as the gauge group. We will use only adjoint representations for
GUT scale fields, and we do not want to add any explicit mass terms. In
this case our superpotential will be limited to terms involving three adjoints per term, 
that is, terms of the form
\beq
\lambda_{i j k} A_i A_j A_k ,
\eeq
for a general set of adjoint fields $A_i$.
In general, if the fields are allowed to acquire VEVs proportional to the generator of the
$U(1)$ hypercharge subgroup
\beq
\langle A_i \rangle = a_i \left ( \matrix{ -{{\rm I_{3 \times 3}} \over 3} & 0  \cr 0 &  {{\rm I_{2 \times 2}} \over 2}  }\right ),
\eeq
we find that the conditions for a zero potential, setting
the auxiliary $F^\dagger$ fields equal to zero, are
generally equal in number to the number of fields. It is 
in principle possible to choose coefficients of the terms in the superpotential
in such a way that one or more of the conditions is redundant, and in that case
there would be a free parameter describing the space of VEVs, that is, there would
be a flat direction. It would not be natural to expect such a special set of coefficients.
Furthermore, one could imagine making one or more condition trivial, that is
\beq
F_i^\dagger = 0 ,
\eeq
thereby removing one of the conditions, however, in this case
the corresponding multiplet remains massless.
Thus, with the general number of conditions to 
minimize the potential equal to the number of parameters, all of the VEVs are completely determined and 
there are no flat directions in a general potential that leaves  no massless modes. 

We will find that with a product group, and the addition of bifundamental representations,
many of the flatness conditions will be trivial for appropriately parameterized VEVs, and so
the VEVs will not be so highly constrained, and flat directions will be possible, as we shall
see an explicit example of in the next section.

\section{An $SU(5) \times SU(5)$ Example}
The ideas of Witten regarding product groups and the requirement of approximate flat directions in the
superpotential using only discrete symmetries have been successfully combined in \cite{dnd}. In this 
section we review one of their models, and in the next we attempt a generalization of this model to 
$SO(10) \times SO(10)$.

A symmetry that is a linear combination of an ordinary discrete symmetry that commutes with $SU(5)$
and a discretized gauge transformation in one of the $SU(5)$'s is what is necessary to split triplets
from doublets in an $SU(5) \times SU(5)$ theory. For a $Z_N$ symmetry with $N$ taken to be odd, an 
appropriate gauge symmetry is
\beq\label{su5sym}
g_1 = \left ( \matrix{\alpha^{-1} &  &  &  &  \cr  &  \alpha^{-1}  & & & &  \cr
  &  & \alpha^{-1} & &  \cr   &  & & \alpha^{{N+3 \over 2}} &  \cr  & & & &
  \alpha^{{N+3 \over 2}}}\right ) ,
\eeq
where $\alpha$ is an $N$'th root of unity. 
One might introduce then as GUT fields representations that transform as bifundamentals under the product group.
In particular one might take the fields $\Phi_i, \bar \Phi_j, i, j = 1, 2$ and take them to transform under 
the ${\bf Z}_N$ symmetry
\beq\label{disym}
\Phi_1 \rightarrow \alpha \Phi_1 , \bar \Phi_1 \rightarrow \alpha^{-1} \bar \Phi_1,
\Phi_2 \rightarrow \alpha^{-{ N + 3 \over 2}} \Phi_2 , 
\bar \Phi_2 \rightarrow \alpha^{ N + 3 \over 2} \bar \Phi_2 .
\eeq 
The combined symmetry, ${\bf Z}_N\prime$, of the discrete gauge transform and ${\bf Z}_N$ is preserved by the following VEVs 
in the bifundamental fields
\beq\label{mesonvev}
\langle\Phi_1\rangle = \langle\bar\Phi_1\rangle =
 \left ( \matrix{v_1 &  &  &  &  \cr  &  v_1  & & &  \cr
  &  & v_1 & &  \cr   &  & & 0 &  \cr  &    & & & 0} \right ),\ \ \
\langle\Phi_2\rangle  = \langle\bar \Phi_2\rangle =
\left(\matrix{0 &  &  &  &  \cr  &  0  & & &  \cr
  &  & 0 & &  \cr   &  & & v_2 &  \cr  &    & & & v_2}\right ) .
\eeq 
There are components of the bifundamental fields that are not eaten by the Higg's
mechanism and need to be made massive by the superpotential without breaking flatness 
or adding any explicit mass parameters. This can be done simply by adding three adjoints
of the first $SU(5)$, $A_{i=1,2,3}$ and a gauge singlet, $S$, with the following potential
\beq\label{su5w}
W = \lambda_{12} \Phi_1 A_1 \bar\Phi_2 + \lambda_{21} \Phi_2 A_2 
\bar \Phi_1 + \lambda_{11} \Phi_1 A_3 \bar \Phi_1 + \lambda_{22} 
\Phi_2 A_3 \bar \Phi_2 + \eta_{12} S A_1 A_2 + \eta_{33} S A_3 A_3 .
\eeq
The gauge singlet in this model gains a VEV $\langle S \rangle = s$, which is another flat
direction of the potential. 

Flatness may be broken by terms that are not forbidden by
symmetry. It is possible to forbid all such terms with a continuous, global symmetry.
However, with discrete symmetries it is at best possible to have approximate flat directions.
Adding a discrete $R$ symmetry, ${\bf Z}_{11}^R$, and assigning charges as in Table~\ref{tab:model} 
one finds that the lowest dimensional flatness breaking terms are
\beq\label{flatbreak}
{1 \over M_{\rm Pl}^6} S^4 (\bar \Phi_1^3 \bar \Phi_2^2), {1 \over M_{\rm Pl}^7} (\Phi_1^3 \Phi_2^2)^2 .
\eeq
These are terms of dimension $9$ and $10$, respectively, and as was argued in Section \ref{doublettriplet}, 
and originally in \cite{dnd}, if the singlet acquires a negative mass-squared in supersymmetry breaking, the
VEVs of these fields will be fixed near $M_{\rm GUT}$.

\begin{table}[ht]
\begin{center}
\begin{tabular}{cc} \hline \hline
Field           &  $SU(5)\times SU(5)\times {\bf Z}_N\times {\bf Z}_{11}^R$ \\ \hline\hline
$\Phi_1$        &  $(5,\bar5,1,0)$      \\
$\bar\Phi_1$    &  $(\bar5,5,N-1,0)$    \\
$\Phi_2$        &  $(5,\bar5,(N-3)/2,3)$        \\
$\bar\Phi_2$    &  $(\bar5,5,(N+3)/2,8)$        \\
$A_1$           &  $(24,1,(N-5)/2,4)$ \\
$A_2$           &  $(24,1,(N+5)/2,9)$ \\
$A_3$           &  $(24,1,0,1)$ \\
$S$             &  $(1,1,0,10)$ \\ 
$h$             &   $(5,1,0,1)$ \\
$\bar{h}'$    &   $(1, \bar{5},0,0)$ \\ \hline\hline
\end{tabular}
\end{center}
\caption{}
\label{tab:model}
\end{table}

Now add the following Higgs fields, $h$ and $\bar{h}'$ that transform under the 
gauge group as
$(5, 1)$ and $(1, \bar{5})$, respectively. 
Also, give $h$ charge one under ${\bf Z}_{11}^R$, and give $\bar{h}'$ $R$-charge zero, as indicated in
the table.
This will then allow the following term
\beq
W_1 = h \bar{\Phi}_1 \bar{h}' .
\eeq
This will then give mass to the triplet fields, and leave the doublets massless. There are some problems with this,
however, because as it stands this theory is not anomaly free. Adding another pair of Higgs fields transforming in the
opposite $SU(5)$ fixes this, but adds an extra massless doublet. One might also imagine cancelling the anomaly 
with standard model matter fields. These issues are discussed in greater detail in \cite{dnd}.

\section{SO(10) Model Building}

We want to consider now to what extent the model described in the
previous section
can be generalized to other groups. Given the success of 
$SO(10)$ unification, $SO(10) \times SO(10)$ is a natural place to start.
We first do this somewhat naively, but will find that the generalization does
not carry over as well as we might hope. 
In $SO(10) \times SO(10)$, much as in $SU(5) \times SU(5)$,
we should first see how it might be possible to implement Witten's ideas for
solving the doublet-triplet splitting with a discrete symmetry. Suppose one has two bifundamentals of
$SO(10) \times SO(10)$, $\Phi_1, \Phi_2$. The unbroken discrete symmetry may be a linear combination of
a discrete symmetry acting on these bifundamentals, and a discrete gauge transformation. 
Taking this gauge transformation to be a discrete hypercharge transformation in one of the
$SO(10)$'s
but not the other will forbid only doublet
masses as prescribed by Witten. The $SU(5)$ subgroup
of a single $SO(10)$ is generated by the following generator of $SO(10)$ in the fundamental
representation, written as a direct product of a $5 \times 5$ space and a $2 \times 2$
\beq\label{su5}
A_5 \otimes I_{2 \times 2} + S_5 \otimes i \sigma_2 ,
\eeq
where $A_5$ is an antisymmetric $5 \times 5$ matrix, and $S_5$ symmetric and traceless,
 $\sigma_2$ is the second Pauli 
matrix and $I_{2 \times 2}$ the identity matrix. Knowing this we can find out how hypercharge 
acts in the $SO(10)$
multiplets. The generator of hypercharge is realized in the above notation by setting:
\beq\label{hyperchrg}
S_5 = \left ( \matrix{-1  &  &  &  &  \cr  &  -1  & & & \cr  &  & -1 & &  \cr
& & & 3/2 &  \cr  & & & &  3/2}\right ) .
\eeq
which generates in $SO(10)$ the group transformation in the fundamental representation:
\beq\label{gtntrdfths}
g_h = \left ( \matrix{\alpha^{-2} &  &  &  &  \cr  &  \alpha^{-2}  & & &  \cr
  &  & \alpha^{-2} & &  \cr   &  & & \alpha^{3} &  \cr  &    & & &
  \alpha^{3}}\right ),
  \eeq
where $\alpha$ is a general two dimensional rotation matrix.
We are allowed to combine a discrete hypercharge transformation with a
${\bf Z}_N$ symmetry for the unbroken symmetry.
The general discrete form of the hypercharge transformation that 
we might wish to use is, taking $N$ to be odd (a similar expression may be derived for even $N$) ,
\beq\label{gauge}
g_1 = \left ( \matrix{\alpha^{-1} &  &  &  &  \cr  &  \alpha^{-1}  & & & &  \cr
  &  & \alpha^{-1} & &  \cr   &  & & \alpha^{{N+3 \over 2}} &  \cr  & & & &
  \alpha^{{N+3 \over 2}}}\right ) ,
 \eeq
where $\alpha$ is now a $2 \times 2$ rotation matrix such that $\alpha^N = 1$. For the product group
$SO(10) \times SO(10)$, the discrete hypercharge trasformation may be in either subgroup, and
without loss of generality we may take it to be a subgroup of the right $SO(10)$.
Now take the bifundamentals to transform under the $Z_N$ symmetry
\beq\label{origzn}
\Phi_1 \rightarrow 1_{5 \times 5} \otimes \alpha \Phi_1,\ \ \
\Phi_2 \rightarrow 1_{5 \times 5} \otimes \alpha^{-{N+3 \over 2}} \Phi_2,
\eeq
where this transformation is in general a ten by ten matrix, written as a direct product as before. Note
that this discrete symmetry does not commute with $SO(10)$, but does commute with the $SU(5)$ subgroup generated 
by eqn. (\ref{su5}).
The symmetry ${\bf Z}_N^\prime$ which is a combination of this discrete symmetry and the hypercharge transformation
in eqn. (\ref{gauge}) will solve the doublet-triplet splitting problem in the manner prescribed by
Witten. This symmetry is respected by the following  VEVs,
written in block diagonal form, of the bifundamentals
\beq\label{bifvev}
\langle\Phi_1\rangle =
  i*v_1 \left( \matrix{\sigma_2 &  &  &  &  \cr  &  \sigma_2 & & &  \cr
  &  & \sigma_2 & &  \cr   &  & & 0 &  \cr  &    & & & 0 } \right ) +
  u_1 \left( \matrix{1 &  &  &  &  \cr  &  1 & & &  \cr
  &  & 1 & &  \cr   &  & & 0 &  \cr  &  & & & 0 } \right ),  
\eeq
$$ \langle\Phi_2\rangle  =
  i*v_2 \left( \matrix{0 &  &  &  &  \cr  &  0  & & &  \cr
  &  & 0 & &  \cr   &  & & \sigma_2 &  \cr  &  & & & \sigma_2 }\right ) +
  u_2 \left( \matrix{0 &  &  &  &  \cr  &  0 & & &  \cr
  &  & 0 & &  \cr   &  & & 1 &  \cr  &  & & & 1 } \right ).
$$
${\bf Z}_N^\prime$ will be unbroken after symmetry breaking occurs, and the above VEVs will be a minimum of the
resulting potential. If we look at the diagonal subgroup of $SO(10) \times SO(10)$, then the above VEVS
are those of an adjoint of the single $SO(10)$ and a traceless symmetric tensor, the $54$. Assuming the
adjoint VEV is non-zero, then the diagonal subgroup is broken down to 
$SU(3) \times SU(2) \times U(1) \times U(1)$, just as is usual in GUT models involving one $SO(10)$
where symmetry breaking is done by an adjoint \cite{Georgi}. If only the symmetric VEVs are present,
then the diagonal symmetry group is broken to $SO(6) \times SO(4)$.
In either case, similar to single $SO(10)$ models, we will need to introduce a rank breaking 
sector to break the diagonal symmetry group to $SU(3) \times SU(2) \times U(1)$.  

We have so far naively assumed that the above VEVs do break the product group to a simple group, and in the
above we therefore only considered the diagonal subgroup as preserved symmetries. 
However, unlike $SU(5) \times SU(5)$ examples, there can be unbroken off-diagonal symmetries, by which 
we mean symmetry transformations that are different group elements of each simple subgroup, for generalized
parameters in the above VEVs. This implies the existence of extra gauge symmetries leftover in models of 
this type, and so the symmetry breaking has not in fact broken the GUT group sufficiently, to the groups
of the Standard Model.
The existence of such off diagonal symmetries may be seen indirectly in the following model
through analysis of the Goldstone modes
\beq\label{toy}
W =  \lambda_1 \Phi_1 S_1 \Phi_1 + \lambda_2 \Phi_2 S_1 \Phi_2 + \lambda_3 \Phi_1 S_2 \Phi_2
+ \lambda_4 \Phi_1 A \Phi_2 + \lambda_5 X A A + \lambda_6 X S_1 S_1 + \lambda_7 X S_2 S_2 .
\label{super}
\eeq
The $\Phi$ fields are the bifundamentals with VEVs as already described. The $X$ field is
a gauge singlet, and in general may acquire a VEV. The $S$'s are symmetric representations
of, say, the first $SO(10)$, and $A$ is in the adjoint representation of the same. 
Although this model may be looked
at as a generalization of the $SU(5) \times SU(5)$ model described earlier, the point of studying 
it here is to see the effects of unbroken off-diagonal symmetries.
The bifundamentals branch into a symmetric plus an adjoint plus a singlet in going from the 
product group to a single $SO(10)$. The adjoints branching out of the bifundamentals as
well as the explicit adjoint field contain the $(3,1)$ and $(\bar{3},1)$ representations under
the $SU(3) \times SU(2)$ subgroup. Putting in the VEVs one can compute the following
mass matrix for these modes
\beq
\left ( \matrix{ 0 & 0 & 0 \cr 0 & 0 & - {1 \over 2} \lambda_4 v_1 \cr
0 & {1 \over 2} \lambda_4 v_1 & 2 \lambda_5 x } \right ) ,
\eeq
which has only one massless mode for nontrivial VEVs.
However, breaking $SO(10) \times SO(10)$ to $SU(3) \times SU(2) \times U(1)$ predicts
two massless Goldstone modes in the above mass matrix, assuming no other unbroken symmetries exist. Clearly,
that is not the case, and there do exist other unbroken symmetries outside of the unbroken symmetry in
the diagonal subgroup. It is interesting to note as well that if $v_1 = 0$ in the bifundamental VEV, the 
correct number of massless modes is once again present. It turns out that for that particular choice, 
and also $v_2 = 0$, there are no off-diagonal symmetries, which we will now see from a more detailed analysis
of the symmetries.

To investigate more precisely what symmetries may be unbroken, consider the 
following VEV of a single bifundamental field, which leaves an unbroken $SU(5) \times U(1)$ in the 
diagonal subgroup:
\beq\label{vvev}
\langle\Phi\rangle =
  i*v \left( \matrix{\sigma_2 &  &  &  &  \cr  &  \sigma_2 & & &  \cr
  &  & \sigma_2 & &  \cr   &  & & \sigma_2 &  \cr  &    & & & \sigma_2 } \right )
 \eeq
there exists a transformation that preserves this VEV but transforms
differently in each $SO(10)$, that is it is not in the diagonal subgroup.
In fact for group elements expressed in the fundamental representation of each group,
they differ only by a sign. Written as a group element in one of the groups this is:
\beq\label{gelement}
g = O_5 \otimes \sigma_1 ,
\eeq
where $O_5$ is any 5 by 5 orthogonal matrix, and the direct product is taken
with the Pauli matrix, and the group element acting on group indices transforming under the second $SO(10)$
has the opposite sign when acting in the fundamental representation. That is
\beq
g = (O_5 \otimes \sigma_1 , - O_5 \otimes \sigma_1)
\eeq
in the bifundamental representation.
A similar group element can be written 
using $\sigma_3$. Thus, there exists a subgroup of $SO(10) \times SO(10)$
that is not broken and is not entirely in the diagonal subgroup.  

In the case of two 
bifundamentals with the VEVs chosen as in eq. (\ref{bifvev})
to try to break to the standard model, basically the same problem occurs for general choices 
of the parameters, as we have 
already seen indirectly from the Goldstone modes in the 
model above in eq. (\ref{toy}). However, we will find that there is one acceptable choice of parameters
for which there are no preserved off-diagonal symmetries. For simplicity, consider again just one bifundamental,
but give it the following VEV:
\beq\label{vev}
\Phi_0 = \langle\Phi\rangle =
  i*v \left( \matrix{\sigma_2 &  &  &  &  \cr  &  \sigma_2 & & &  \cr
  &  & \sigma_2 & &  \cr   &  & & \sigma_2 &  \cr  &    & & & \sigma_2 } \right ) +
 u \left( \matrix{1 &  &  &  &  \cr  &  1 & & &  \cr
  &  & 1 & &  \cr   &  & & 1 &  \cr  &  & & & 1 } \right ).
\eeq
Now we can consider how this transforms under an infinitesimal $SO(10) \times SO(10)$
transformation. In particular, the infinitesimal generators of transformations by the left
$SO(10)$  can be written as
\beq\label{left}
\delta \Phi = [A_2^L \otimes 1_{2 \times 2} + S_2^L \otimes i \sigma_2 + A_1^L \otimes \sigma_1
 + A_3^L \otimes \sigma_3] \Phi
\eeq
and similarly the infinitesimal transformation from the right is
\beq\label{right}
\delta \Phi = \Phi [-A_2^R \otimes 1_{2 \times 2} + S_2^R \otimes i \sigma_2 - A_1^R \otimes \sigma_1
 - A_3^R \otimes \sigma_3].
\eeq
The notation here is that the $A$'s are antisymmetric five by five matrices, and the $S$'s are symmetric.
The infinitesimal transformation is then the direct products of dimension five and two matrices.
The minus signs in the transformation from the right reflect that this is the transpose of 
the fundamental transformation (since we compare group elements in the fundamental representation
to see if we are in a diagonal subgroup, we cannot just take the signs away as a matter
of convention without putting signs in somewhere else). A general transformation that 
combines left and right may be written
in terms of commutators and anti-commutators. Furthermore, we are not interested
in the first two parts of each of the above transformations, because they can only generate
diagonal transformations that are in the diagonal $SU(5) \times U(1)$ subgroup.
The combined transformation with only the $\sigma_1$ and $\sigma_3$ parts can be written
as follows
\beq\label{combined}
\delta \Phi = [A_1 \otimes \sigma_1 , \Phi ] + \{ \tilde{A_1} \otimes \sigma_1 , \Phi \}
 +  [A_3 \otimes \sigma_3 , \Phi ] + \{ \tilde{A_3} \otimes \sigma_3 , \Phi \},
\eeq
where the five by five antisymmetric matrices above are linear functions of $A_1^R$, $A_1^L$, 
$A_3^R$, and $A_3^L$ in the original left and right transformations. 
For $A_1 = ({ 2 u \over v }) \tilde{A_3}$ and
$A_3 = - ({ 2 u \over v} ) \tilde{A_1}$ this preserves the above VEV: 
$\langle \Phi \rangle = \Phi_0$. We may rewrite this as follows
\beq\label{inftrans}
\delta \Phi = ( 2 u [ A \otimes \sigma_1 , \Phi] + v \{ A \otimes \sigma_3 , \Phi \} ) +
 ( 2 u [\tilde{A} \otimes \sigma_3 , \Phi] - v \{\tilde{A} \otimes \sigma_1 , \Phi \} ) .
\eeq
The transformation generated above is in general not in the diagonal subgroup of the
dual group, with the exception that if $v = 0$ this transformation is just $A_1^L=A_1^R$ 
and $A_3^L=A_3^R$ in the notation of the fundamental, dual group.
This latter case corresponds to merely having broken $SO(10) \times SO(10)$
to a single, diagonal $SO(10)$. If we now change our VEV to the original form in eq. (\ref{bifvev})
and follow the same procedure as above we find that if $v_1=v_2=0$ there are no off diagonal symmetries
and the preserved diagonal subgroup is $SO(6) \times SO(4) \times U(1)$. 
This situation is somewhat analogous to what happens in single $SO(10)$ GUTs when a field
transforming as a symmetric (54) is used to break the symmetry \cite{Georgi}. Just as in the case of breaking
with an adjoint field, an additional rank breaking sector is required when the GUT breaking field
is a symmetric (54). 
A rank breaking sector will also work for dual group models because the
dual symmetry is already broken to a diagonal subgroup.

\section{Rank Breaking and Flatness}
In single $SO(10)$ 
models a $16 + \ov{16}$ or $126 +\ov{126}$ sector can be used to break the rank of $SO(10)$ 
\cite{Georgi}. As was found above, a rank breaking sector consisting of the appropriate fields that 
transform under either group is what is necessary.
For our models we require a flat potential for the fields that
break the rank of $SO(10)$. Specifically, we require that no VEVs are fixed and no GUT scale masses are
put in as input parameters. We need to satisfy these requirements, or do so approximately with possible
nonrenormalizable terms breaking flatness as in the section above, to allow for a natural
way of generating the GUT scale. We also need to make sure that all the fields are
massive. It will be found, however, that it is impossible to construct a rank breaking sector
satisfying these criteria. One other requirement that should be mentioned is that the rank 
breaking sector must couple to the rest of the theory involving bifundamentals, 
without spoiling their VEVS, in order to avoid light pseudogoldstone bosons.

To start off assume we use $16 + \ov{16}$ fields to break the rank, and we will see
that the same issues will carry over to the $126 + \ov{126}$ case. The programme we are
to follow then is to try to invent terms and add fields as needed to try to give mass to
all the fields without fixing the $SU(5)$ singlet VEVs of the 16s or using explicit mass terms
(e.g. all masses come from VEVs, the fundamental theory has no mass terms). Specifically,
we might look at the 16 field's branching rules into $SU(3) \times SU(2)$ from the standard 
model, without worrying about hypercharge. 
\beq\label{counting}
16 = \bar 5 + 10 + 1 = ( \bar 3 , 1 ) + ( 1 , 2 ) + ( 3 , 1 ) + ( 3 , 2 )
 + {\rm Singlets}
\eeq
Here we have first written the $SU(5)$ fields and then the branching rules for SM representations. 
The doublet fields are of particular interest, for the simple reason 
that making them massive is very difficult under the 
requirement of flat potentials
and no explicit mass terms. Furthermore, in the Standard Model there are no
doublets eaten by the Higgs mechanism, so it is a requirement that these doublets
acquire masses from the VEVs.

Firstly, there does not appear to exist another representation of $SO(10)$ that contains 
a standard model doublet and can couple to the $16 + \ov{16}$ . This assertion can be
verified up to well studied representations using tables of branching rules, such as those in \cite{Slansky}. The 
Standard Model doublets generally branch out of the $5$ and $\ov{5}$ representations
of $SU(5)$. The $120$ of $SO(10)$, for example, contains both a $5$ and a $\ov{5}$, but
it is a three index antisymmetric tensor, and by itself cannot couple to the $16 + \ov{16}$. 
One can look at some higher representations and find similar results.
It is rather difficult to find the right field because we are required to use
a field that couples to both $16$ and $\ov{16}$ and we are restricted to
products of three fields in the superpotential. In any case, if we restrict ourselves
to representations commonly used in $SO(10)$ model building, in particular the 
adjoint and symmetric tensor, then we know there is no field with doublets that we
can couple to the $16 + \ov{16}$.
As a result, the only way to make the doublets massive is to couple the 16's to some other field
that acquires a VEV, and then hope the VEV generates sufficient mass terms. 
The 16 fields' $F$-fields equal to zero require, however, that there is more than one coupling to these fields,
because were there only one coupling, the $F$-fields require that the field coupled to has zero 
VEV. That is to say, a coupling of the form $\bar C A C$ with no other fields coupling to the
$16 + \bar 16$ fields implies that $A$ has zero VEV, or at the very least no VEV that could give mass
to doublets. 

The only remaining possibility is to add another field to couple to the $16$s. This
field is necessarily in a different representation of the symmetry group because otherwise
we could redefine the fields in such a way that there is only one coupling, again.
However, we will find that in general this will fix all the VEVs, in particular the 
$16$s' VEVs are constrained by the $F$-fields for each field they couple to. With two such 
constraints, the $16$ VEV is fixed except for very special coincidences of coupling constants
which could not be justified in a natural way, in the sense that there is no symmetry to justify
such special choices of coeeficients. To see this, consider just two fields coupling
to the $16$'s, and these two couple only to each other. Then there are two constraints 
coming from setting the $F$'s equal to zero for the two added fields, and the constraint
coming from the $16$'s, which is generally constrained. Adding fields adds constraints 
and additional VEV parameters in equal number, generally, so the VEVs have fixed values in
general.

The $126$ also contains a single Standard Model doublet, and the issues with coupling to 
other fields are similar to the $16$'s, so the same arguments generally hold for the $126$
rank breaking fields.

One might also ask whether it might be possible to introduce some generalization
of bifundamental fields, that is, fields that transform under both groups, but generalizing
this beyond the fundamental representation of each group. The problem is, we still need
the bifundamental fields we introduced before to break to $SO(6) \times SO(4) \times U(1)$.
Once that is done all the fields transform in the diagonal subgroup, and we are left with
essentially the same predicament.

\section{Conclusions}
With the requirement of a flat potential so that $M_{\rm GUT}$ is a modulus 
we have shown that models with
$SO(10) \times SO(10)$ gauge group are impossible 
to construct. We also have shown that
models with a single $SU(5)$ gauge group without input mass terms do not allow flat potentials, 
and this appears to be general to simple groups without exotic representations in the theory. 
This means that 
the GUT masses are still functions of parameters we put in by hand in theories with a simple
gauge group. Product groups
provide a natural way to solve the problem of splitting doublets from triplets, as well as in the 
case of $SU(5) \times SU(5)$ providing a way to break the unification group at the GUT scale 
that leaves the scale of this breaking a modulus. What we have essentially shown is that this set of 
ideas 
used successfully for $SU(5) \times SU(5)$ cannot be generalized to $SO(10) \times SO(10)$, nor groups 
containing $SO(10) \times SO(10)$ because ultimately these kinds of models would run into the same problems
with rank breaking.

\end{document}